# Are Scientists Changing their Research Productivity Classes When They Move Up the Academic Ladder?

Marek Kwiek[1,2] · Wojciech Roszka[1,3]

**Abstract**
We approach productivity in science in a longitudinal fashion: We track scientists' careers over time, up to 40 years. We first allocate scientists to decile-based publishing productivity classes, from the bottom 10% to the top 10%. Then, we seek patterns of mobility between the classes in two career stages: assistant professorship and associate professorship. Our findings confirm that radically changing publishing productivity levels (upward or downward) almost never happens. Scientists with a very weak past track record in publications emerge as having marginal chances of becoming scientists with a very strong future track record across all science, technology, engineering, mathematics, and medicine (STEMM) fields. Hence, our research shows a long-term character of careers in science, with one's publishing productivity during the apprenticeship period of assistant professorship heavily influencing productivity during the more independent period of associate professorship. We use individual-level microdata on academic careers (from a national registry of scientists) and individual-level metadata on publications (from the Scopus raw dataset). Polish associate professors tend to be stuck in their productivity classes for years: High performers tend to remain high performers, and low performers tend to remain low performers over their careers. Logistic regression analysis powerfully supports our two-dimensional results. We examine all internationally visible Polish associate professors in five fields of science in STEMM fields (N=4,165 with $N_{art}$=71,841 articles).

✉ Marek Kwiek
  kwiekm@amu.edu.pl

  Wojciech Roszka
  wojciech.roszka@ue.poznan.pl

1   Center for Public Policy Studies, Adam Mickiewicz University, Poznan, Poland
2   German Center for Higher Education Research and Science Studies (DZHW), Berlin, Germany
3   Poznan University of Economics and Business, Poznan, Poland

## Introduction

In the present article, we address a simple research question about the impact of prior individual research productivity on current research productivity. Assuming that scientists can change productivity classes during their careers, we seek patterns of mobility between productivity classes in the five fields of science comprising 12 disciplines.

We examine the changing productivity of 4,165 Polish science, technology, engineering, mathematics, and medicine (STEMM) scientists as they move up the academic ladder. All examined scientists are associate professors employed full time in the higher education sector, and they all have both doctoral and postdoctoral (habilitation) degrees. Combining demographic and biographical data based on a national registry of scientists (N=99,935) with our own computations based on Scopus metadata on all Polish research articles indexed over the past half a century (1973–2021, N=935,167), we examine individual scientists who change their productivity classes over time, here for a period spanning up to 40 years (range of biological age in the sample: 30–70). Our focus is on the two career stages of assistant professorship and associate professorship, at which the vast majority of Polish academic scientists are currently located (GUS, 2023).

Our point of departure is allocating all associate professors who are internationally visible in the Scopus database to 10 current productivity classes for the period of 2018–2021 (by productivity deciles). Then, we examine their past productivity when they were assistant professors, compare them with their peers in their own fields of science, and retrospectively allocate them to 10 past productivity classes (again from the top to the bottom classes). We unpack the details of scientists' individual trajectories in these two career stages, linking current and past productivity for each individual scientist, before then examining the mobility between productivity classes by field of science and productivity type. In particular, we are interested in comparing mobility patterns between productivity classes in terms of four types of productivity—full counting and fractional counting—in both (journal) prestige–normalized and non-normalized versions.

Consequently, our approach is longitudinal (tracing the productivity of the very same scientists over time) and classificatory (examining productivity changes in terms of 10 decile-based productivity classes rather than publication numbers; the top 10% is a classic measure of productivity inequality, see, e.g., the "top scientists" in Abramo et al., 2017). We examine scientists from the top and from the bottom productivity classes who change classes over time from a relative perspective: Class identification is possible by studying the productivity of individuals in relation to the productivity of other individuals (as in the studies of research stars; see, e.g., Aguinis & O'Boyle, 2014: 313–315; DiPrete & Eirich, 2006: 282).

## Theoretical Framework

### Persistent Inequality in Academic Knowledge Production

Based on prior research on high productivity (Abramo et al., 2009a, 2009b; Fox & Nikivincze, 2021; Kwiek, 2016; Yin & Zhi, 2017), we focus on the persistence of top productivity and of bottom productivity over time as scientists move up the academic ladder. Our intuitions are based on prior theories in the sociology of science and economics of science, according to which top-productive scientists tend to keep being top-productive and bottom-productive scientists tend to keep being bottom productive while nonproductive scientists tend to leave the academic science sector (Allison & Stewart, 1974: 596; Allison et al., 1982: 615; Cole & Cole, 1973: 114; Turner & Mairesse, 2005: 3).

The steep performance stratification of scientists and persistent inequality in academic knowledge production have been examined for a long time, with foundational analyses being given by Alfred Lotka (1926), de Solla Price (1963), Robert K. Merton (1968), Cole and Cole (1973) and others inspiring generations of theoreticians. The old research theme, summarized as "the majority of scientific work is performed by a relatively small number of scientists" (Crane, 1965: 714), has been at the core of these theories of individual research productivity.

The mechanisms behind accumulative advantage (and disadvantage) have been studied for decades (Alison et al., 1982; Allison & Stewart, 1974; Cole & Cole, 1973; DiPrete & Eirich, 2006; Merton, 1968), as have other major theories of research productivity, such as sacred spark theory (Allison & Stewart, 1974; Cole & Cole, 1973; Fox, 1983; Zuckerman, 1970) and utility maximization theory (Kyvik, 1990; Stephan & Levin, 1992). Built-in undemocracy seems to be part and parcel of research performance, and "inequality has been, and will always be, an intrinsic feature of science" (Xie, 2014: 809). In Poland, as elsewhere, low productive scientists work in STEMM laboratories alongside highly productive scientists (Abramo et al., 2013; Piro et al., 2016)—and in Poland, 10% of the most productive scientists ("research top performers") have been shown to be producing as much as 50% of all publications (Kwiek, 2018). The role of research stars, who are concentrated in the right tail of research productivity distribution in every national science system, has endured over time (Agrawal et al., 2017: 1). The skewness of science has been the topic of numerous bibliometric publications (e.g., Albarrán et al., 2011; Carrasco & Ruiz-Castillo, 2014; Ruiz-Castillo & Costas, 2014). Recent studies include research on variously termed highly productive scientists: stars and superstars (Abramo et al., 2009a, 2009b; Agrawal et al., 2017; Aguinis & O'Boyle, 2014; Sidiropoulos et al., 2016; Yair et al., 2017), the best (O'Boyle & Aguinis, 2012), prolific professors (Piro et al., 2016), top researchers (Abramo et al., 2013; Cortés et al., 2016), and the academic elite (Kwiek, 2016; Yin & Zhi, 2017).

### Research Productivity

In most science systems, research productivity is one of the most important dimensions—although not the only one—determining the trajectory of academic careers (Leišyte & Dee, 2012; Stephan, 2015). Research productivity has been widely studied from both single-nation and cross-national perspectives (see, e.g., Allison et al., 1982; Fox, 1983; Kwiek, 2018; Lee & Bozeman, 2005; Ramsden, 1994; Shin & Cummings, 2010; Stephan & Levin, 1992; Teodorescu, 2000; Wanner et al., 1981). In addition to publications, a successful academic career is determined by factors such as external research funding obtained, patterns of international collaboration, awards and honors, membership in associations and academies, physical mobility and international experience, professional networks, institutional placement (i.e., institutional and national affiliation), and luck (Carvalho, 2017; Hermanowicz, 2012).

Career success is also determined by the internationalization of research, citations received, working time distribution, distribution of academic roles, and other factors. The main drivers behind productivity fall into two types: individual and environmental (encompassing both institutions, in the form of, e.g., "work climate," as shown by Fox & Mohapatra, 2007, and entire national science systems, in the form of, e.g., national academic promotion and recognition systems, as shown by Leišyte & Dee, 2012).

Within the most general, traditional tripartite division of academic tasks into teaching, research, and service, it is extremely difficult to compare researchers' achievements in the first and third areas, mainly because of data limitations. In contrast, it is relatively more simple, though not without controversies, to compare achievements in the area of research through publications, which are usually indexed in global databases and their citations. Because publication and citation databases (despite their limitations and biases, as widely discussed in the literature; see Baas et al., 2020; Sugimoto & Larivière, 2018) have metadata of publications spanning for decades, it is possible to analyze individual productivity (calculated as the number of publications of a selected type per unit of time) and how it changes over time.

However, studying changes in productivity over time requires the data at the individual scientist level rather than at the publication level, which, in turn, requires massive processing of publication-oriented bibliometric data into a different unit of analysis: the individual scientist. In addition, studying productivity changes over time using publication numbers faces additional limitations because of the different pace of development of bibliometric databases, here depending on the discipline.

In some disciplines, the increasing number of publications may be because of increasing individual productivity, while in others, it may be because of the increasing number of journals successively included in the database. Moreover, the average productivity increases at different rates in different disciplines with successive generations of scientists—scientists not only start publishing earlier on average, but they also publish more per year on average (Wang & Barabási, 2021). Higher productivity has also been associated with the growing role of multiauthored and internationally coauthored publications and the increasing average size of research teams (Adams, 2013; Wuchty et al., 2007), which, in turn, have been associated with increasing specialization in science and a stronger imperative to show the contributions of all, even minor, participants in research.

**Related Research and Research Gaps**

Our longitudinal and classificatory approach to individual research productivity is especially promising for systems in which digital biographical and demographic data on scientists (from national registries) are available. In the literature, there are at least three somewhat structurally similar studies: These studies have focused on a national system (Abramo et al., 2017 on Italy), a single institution (Kelchtermans & Veugelers, 2013 on KU Leuven in Belgium), and a single discipline in a country (Turner & Mairesse, 2005 on French condensed-matter physicists). These authors explored persistence of research productivity over time, with varying periods of time and different datasets being examined (a national ministerial dataset, institutional personnel administrative data, and bibliometric data).

Abramo et al. (2017) studied the persistence of "stardom" of scientists (or their membership in the upper 10% in terms of productivity), focusing on the top performance of all Italian professors over three 4-year periods (2001–2012). They identified the top performers in the first period (N=2,883) and tracked them over time in the next two periods. The authors showed that about one-third of top performers retain their stardom for three consecutive periods, and about half retain it for two periods (35% and 55%, respectively, with some disciplinary differentiation and with higher percentages for male scientists, Abramo et al., 2017: 793–794).

In their study of KU Leuven, Kelchtermans and Veugelers (2013) examined persistence of research productivity over time at an individual level using a panel dataset comprising the publications of 1,040 biomedical and exact scientists for the period 1992–2001. They examined how researchers switch between the three productivity categories (top, medium, and low classes) over time and showed that productivity categories are generally persistent over time. Next top performance was found to be positively affected by previous top performance.

Finally, for 497 French physicists in the periods of 1986–1991 and 1992–1997, Turner and Mairesse showed that 66% of the most productive researchers (defined as "quartile 1 scientists") and 67% of the least productive researchers (defined as "quartile 4 scientists") remained as such for the entire period 1986–1997, underlying a stability of the relative positions of the researchers in the distribution of publication counts over time.

Our approach is different in several respects: the direction of tracking scientists over time (retrospective tracking of individuals vs. forward tracking); the period covered (two career periods, assistant professorship and associate professorship, spanning up to 40 years); the construction of the sample (all internationally visible associate professors within a national system); and the methodology (analysis of 10 decile-based productivity classes, from the top 10% to the bottom 10%; and the four approaches to productivity, including two journal prestige normalized). Finally, we have used logistic regression analysis to identify major predictors of membership in the top and bottom productivity classes. However, in very general terms, Abramo et al.'s (2017) analysis of the "stardom" of scientists over time within a national population of scientists bears the most interesting similarities with our analysis of the mobility between top productivity classes and bottom productivity classes over time as individuals move up the academic ladder; Kelchtermans and Veugelers (2013) and Turner and Mairesse (2005) used different methodological approaches that are not directly comparable to ours. Specifically, none of the above approaches examined the top-to-bottom and bottom-to-top (rather than merely top-to-top and bottom-to-bottom) mobilities between productivity classes.

The literature shows several important gaps which we intend to fill: first, the vast majority of productivity literature is based on cross-sectional (mostly survey-based) rather than longitudinal datasets. Second, those few longitudinal studies are focused on top-to-top mobility in productivity which does not reflect patterns in a plethora of individual careers in which scientists move up or down in their (field-normalized) productivity; from an individual perspective, extremely rare bottom-to-top mobility is as important as much more frequent top-to-top mobility from an institutional perspective. Third, the literature has not shown persistence in productivity over time in a more granular manner (e.g., all productivity deciles can be studied, with the majority of mobility to the top deciles coming from the neighboring deciles and with no mobility coming from the bottom deciles); more general mobility patterns – such as scientists moving between productivity quartiles – hide behind them more nuanced patterns for which a more detailed approach is needed. A focus on the mobility of the top 25% of scientists reveals different patterns than a focus on the top 10%. Additionally, with our micro-level data, we are able to go down to individuals with their IDs, with their unique publishing and collaboration profiles and their idiosyncratic career, promotion, and tenure details. Fourth, research has not explored the issue of the impact of different counting methods on the scale of mobility observed; specifically, the role of the vertical hierarchical structure of the academic journal system (and the role of differentiated and citation-based, measurable journal prestige) has not been taken into account. Different counting methods play an extremely powerful role in those science systems – as in Poland – in which grant allocation, promotions, and tenure decisions are strictly publication-related. Standard productivity (with no journal prestige-normalization) seems powerless when applied to real-life productivity in which publications in some outlets – officially defined by the ministry of science and unofficially recognized by the scientific community – matter for individuals and institutions and publications in other outlets do not matter at all. Finally, current econometric models used in longitudinal studies do not employ the registry-based biographical data so that, for example, the promotion speed classes and promotion age classes cannot be used in explaining high productivity in the way both are used in our research.

We have previously examined the persistence of productivity classes in the top levels of Polish academia (i.e., among full professors) from a lifetime perspective using a different methodology (Kwiek & Roszka, 2024). For a sample of 2,326 full professors from 14 STEMM disciplines, we have previously observed several consistent productivity patterns. Using the 20/60/20 classification (as opposed to the current more fine-grained approach: 10 productivity deciles) into top performers, middle performers, and bottom performers, we analyzed current full professors retrospectively. We have shown that, for the sample of current full professors, half of the highly productive assistant professors in the past continued to become highly productive associate professors, while half of the highly productive associate professors continued to become highly productive full professors (52.6% and 50.8%, respectively). In logistic regression models, there were two powerful predictors of membership in the top productivity class for full professors: first, being highly productive as assistant professors and, second, being highly productive as associate professors earlier in their careers.

## Dataset, Sample, and Methodology

### Dataset

For the present research, two major data sources are combined: national and international. The national dataset is the "Polish Science Observatory" dataset, which has been created and is maintained by the present authors. The international dataset is Scopus raw publication and citation data for 1973–2021 for all Polish scientists and scholars who are active in performing research during the period. The "Observatory" database has been created by merging a national biographical and administrative register of all Polish scientists and scholars (N = 99,935) with the Scopus bibliometric database (2009–2018, metadata on N = 380,000 publications of authors with Polish affiliations). The Observatory includes relevant data such as gender, date of birth, dates of academic promotions (doctoral degree, postdoctoral degree, professorship title, if applicable), present institutional affiliations, and disciplines in which degrees have been obtained.

The official national register and the 2009–2018 Scopus publication and citation database have been merged using probabilistic and deterministic methods (see Kwiek & Roszka, 2021: 4–6). The Observatory database has been subsequently enriched with publication metadata for all scientists and scholars with Polish affiliations for the past half century collected from Scopus and obtained through a multiyear collaborative agreement with the International Center for the Studies of Research (ICSR) Lab, a cloud-computing platform provided by Elsevier for research purposes (N = 935,167 articles from 1973–2021).

### Sample

Our sample (N = 4,165 scientists with $N_{art}$ = 71,841 articles) includes scientists currently employed full time in higher education institutions at the rank of associate professors and who have both doctoral and postdoctoral (habilitation) degrees and are working in one of the five STEMM fields of science (composed of 12 STEMM disciplines; their list is provided in Table 1).

Our sample is about one-third female scientists and two-thirds male scientists (37.3% and 62.7%, respectively), generally reflecting the gender structure of the Polish academic profession in STEMM fields of science at the rank of assistant and associate professors. As elsewhere, the share of female scientists in Poland is the highest for lower ranks and the lowest for higher ranks, reaching 28.34% of women working at the rank of full professors for all STEMM and non-STEMM disciplines combined (GUS, 2023: Table 1/42).

Table 1 Structure of the sample of all Polish internationally visible associate professors by gender, age group, and STEMM academic field (N=4,165) (frequencies and percentages)

|  |  | Total | | Female scientists | | | Male scientists | | |
| --- | --- | --- | --- | --- | --- | --- | --- | --- | --- |
|  |  | n | col % | n | col % | row % | n | col % | row % |
| Age groups | Total | 4165 | 100.0 | 1553 | 100.0 | 37.3 | 2612 | 100.0 | 62.7 |
|  | 39 and less | 301 | 7.2 | 92 | 5.9 | 30.6 | 209 | 8.0 | 69.4 |
|  | 40–54 | 2575 | 61.8 | 1036 | 66.7 | 40.2 | 1539 | 58.9 | 59.8 |
|  | 55 and more | 1289 | 30.9 | 425 | 27.4 | 33.0 | 864 | 33.1 | 67.0 |
| IDUB | IDUB | 1240 | 29.8 | 354 | 22.8 | 28.5 | 886 | 33.9 | 71.5 |
|  | Rest | 2925 | 70.2 | 1199 | 77.2 | 41.0 | 1726 | 66.1 | 59.0 |
| Field | ENGI | 959 | 23.0 | 184 | 11.8 | 19.2 | 775 | 29.7 | 80.8 |
|  | LIFE | 897 | 21.5 | 485 | 31.2 | 54.1 | 412 | 15.8 | 45.9 |
|  | MATH | 400 | 9.6 | 76 | 4.9 | 19.0 | 324 | 12.4 | 81.0 |
|  | MED | 630 | 15.1 | 335 | 21.6 | 53.2 | 295 | 11.3 | 46.8 |
|  | NATURAL | 1279 | 30.7 | 473 | 30.5 | 37.0 | 806 | 30.9 | 63.0 |

Twelve Scopus ASJC disciplines (All Science Journal Classification) from STEMM have been clustered into five fields of science. The fields of science included are the following: ENGI (Engineering, composed of Engineering and Materials Science); LIFE (Life Sciences, composed of Agricultural and Biological Sciences; and Biochemistry, Genetics, and Molecular Biology); MATH (Mathematics, composed of Mathematics and Computer Science); MED (Medicine, composed of Medical Sciences); and NATURAL (Natural Sciences, composed of Chemical Engineering, Chemistry, Physics and Astronomy, Earth and Planetary Sciences, and Environmental Science)

Almost two-thirds of associate professors in our sample are aged 40–54 (61.8%), and both male and female associate professors are scattered across the three age groups, with only less than one-tenth aged under 40 (7.2%). The Kernel density plot in Fig. 1 indicates that the current age distribution of associate professors by gender differs, especially for older age groups. Specifically, the share of older age group associate professors is higher for men than women, which may reflect a higher inflow of women to STEMM disciplines 30 years ago and earlier. One in three scientists (29.8%) come from research-intensive institutions defined as IDUB institutions, that is, the participants in the first Polish national excellence initiative (funded with an additional 1 billion USD for 2020–2026). The age distribution of scientists by fields of study is only slightly differentiated, as the Kernel density plots indicate: Although in two fields younger age groups dominate (e.g., NATURAL and LIFE), in MATH, scientists are distributed in a much flatter manner, with ENGI having larger shares of older scientists.

## Methodology

### Unit of Analysis: Individual Scientists Rather than Individual Publications

Individual scientists with unambiguously defined biographical and publication-related attributes are the unit of analysis. Scientists have individual identification

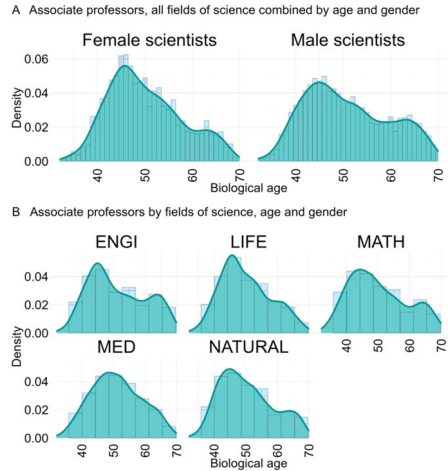

**Fig. 1** Distribution of biological age: Kernel density plots. A. associate professors in five STEMM academic fields combined by gender. B. associate professors by STEMM academic field (N = 4,165)

numbers (IDs), so their career-related biographical data can be derived from a national registry of scientists. The focus is on the individual academic careers of research producers developing over time, in this case on changing productivity classes while moving up the academic ladder, rather than on the research products (i.e., publications) themselves.

### Constructing Individual Publication Portfolios

For each scientist, a unique individual publication portfolio is constructed. The portfolio includes all Scopus-derived metadata on their publications. Specifically, publication metadata are journal metadata (e.g., Scopus CiteScore percentile rank) and publication metadata (e.g., year of publication, number of coauthors and their affiliations, citation numbers). Each publication is linked to the dates in individual biographical history; that is, they are linked to two career steps: assistant professorship and associate professorship, which is clearly defined as the period between obtaining a doctoral degree and a habilitation degree and the period following the conferral of a habilitation degree, respectively. The date of the first publication (any type) in the Scopus database allows for the construction of publication-derived academic age, a proxy of academic experience, which has been used in logistic regression models (we analyzed the correlation between biological age and academic age in Poland in Kwiek & Roszka, 2022).

### Constructing Individual Biographical Histories

For each scientist, apart from a unique individual publication portfolio, an individual biographical history is also constructed, here with the relevant dates: the date of birth (which allows us to infer biological age at the beginning of both career stages: assistant professorship and associate professorship), the date of obtaining a doctoral degree, and the date of obtaining a postdoctoral (or habilitation) degree. All scientists in our sample are associate professors—but also all of them have been assistant professors earlier in their careers. The first stage of their academic career started when they were awarded their doctoral degrees and the second stage when they were awarded their habilitation degree. For both degrees, we have full administrative data, including the date of degree conferral, dissertation title, employing institution and city, defense institution and city, academic discipline, and academic field. The data, which come from a manually curated national registry of scientists, can be treated as fully reliable.

For the purposes of international comparability, we use holding a doctoral degree as a proxy of assistant professorship and holding a postdoctoral degree as a proxy of associate professorship. Assistant professorships are tenure-track positions, and tenure is granted to associate professors (upon obtaining the habilitation degree), with long-term job contracts and job stability for the vast majority of academics. Associate professorships offer higher salaries and greater participation in university self-governance.

**Longitudinal Approach to Studying Academic Careers**

For our analyses, we have chosen all current internationally visible associate professors (i.e., with at least one journal article indexed in the Scopus database), and we look back at their professional careers: We examine their current publishing behavior in a four-year period of 2018–2021 and their past publishing behavior when they have been assistant professors in four-year equivalents in the past.

In traditional longitudinal research designs, the same individuals are followed over time with selected points in time to enable comparative research (Menard, 2002; Singer & Willett, 2003). In following academic careers, a longitudinal design has not been used for technical and cost-related reasons; however, cohort-based approaches with bibliometric data have been recently used (see, e.g., Huang et al., 2020; Milojevic et al., 2018; Wang & Barabási, 2021). Our combination of individual biographical histories (professional life data) with individual publication portfolios (publication and citation data) allows us to produce a retrospective view in which a sizable group of scientists is traced back for several decades in terms of their publishing behavior. This longitudinal approach opens new possibilities for studying academic careers over time.

The application of four major dimensions to examine biographical and bibliometric data becomes possible: gender, age, field of science, and, most importantly, time. Instead of time-limited snapshot views, using a series of cross-sectional accounts, longitudinal analyses that can focus on the change of the scientific workforce over time by various dimensions (e.g., research productivity) is possible.

**Defining Gender, Biological Age, Academic Age, and Fields of Science**

In our sample, all scientists have unambiguously defined gender (a binary approach used in the national registry: male or female) and year of birth. Hence, their biological age at any point in their professional careers is easy to calculate. Their academic age—or the number of years since the first Scopus-indexed publication (any type) used in the logistic regression models—has been collected using an application programming interface (API) protocol.

We use individual publication portfolios (all Scopus-indexed publications lifetime) to determine the dominant discipline: the modal value of ASJC (All Science Journal Classification) disciplines used in Scopus for each scientist. We link all publications (journal articles and chapters in conference proceedings only; henceforth referred to as articles) in the portfolios to ASJC disciplines, and if there are two or more values with the same high occurrence in the portfolio, the discipline is randomly selected from among them. We then cluster 12 disciplines into five fields of science to have higher representation of men and women scientists and to avoid low numbers of observations in some disciplines.

**Measuring Individual Publishing Productivity**

We measure productivity in the four-year reference period of 2018–2021 (termed "current productivity of associate professors") and in earlier periods of their assistant professorship (termed "past productivity of associate professors when they were assistant professors") by using publication data (journal articles) from individual publication portfolios. We need exact dates from the national registry to determine when current associate professors were working as assistant professors, so we have allocated publications from the 2018–2021 reference period and from the assistant professorship period, which have varying lengths, for each individual scientist. Four-year productivity is used in both cases.

**Journal Prestige–Normalized Approach to Publishing Productivity**

We find it reasonable to take an approach to productivity change over time in which we use productivity classes within disciplines (see Costas & Bordons, 2005, 2007) rather than publication numbers. Our journal prestige–normalized approach to productivity locates scientific articles within a highly stratified global structure of journals by taking into account the fact that those articles published in high-prestige journals require, on average, more scholarly effort than articles published in low-prestige journals. The prestige of a journal (here: as expressed as percentile ranks in the Scopus database) is an important element of individual productivity, especially in systems—as in the Polish case (Antonowicz et al., 2021)—in which both the quantity and quality of articles captured by a proxy of journal prestige count toward academic promotion. The basic rule of Polish research assessment exercises termed institutional research evaluations in the past two decades is as follows: "Academic journals are not equal": papers in different journals are given different numbers of points (in the range of 20–200).

Therefore, having individual publication portfolios for each scientist in our sample, we use four approaches to productivity: two journal prestige normalized and two non-normalized. An especially interesting approach is journal

prestige–normalized approach to productivity, which we have developed over the past few years and in which articles are linked to the Scopus journals in which they were published (see Kwiek & Roszka, 2023; Kwiek & Szymula, 2024a). All Scopus journals (N = 46,702 in 2024) have their distinct locations in the Scopus CiteScore percentile ranks, in the range of 0–99, with more prestigious journals generally located in the 90–99 percentile ranks within their disciplines.

While in our non-normalized approach the value of an article in calculating productivity is 1 (using a full counting methodology) (see Waltman & van Eck, 2019), in our prestige-normalized approach, the value will be normalized to journal percentile ranks. In practice, our fine-grained approach applies the idea that the difference between journal prestige, as measured by Scopus, can be better captured by an exponential function rather than a linear function. The exponential function we apply is a mathematical function denoted by $y = x^{2.5}$ (as opposed to a linear function denoted by $y = x$, where x is Scopus CiteScore percentile rank). We have experimented with several exponential functions, and the exponent 2.5 (contrasted with 1.5, 2 and 3) seems to well capture our intentions: It increases the value of articles in highly ranked journals, especially in the 95th to 99th CiteScore percentile ranks, at the expense of the value of articles in bottom-ranked journals, especially below the 50th CiteScore percentile rank (see the formula in Electronic Supplementary Material).

Linking articles to their location in a highly stratified system of academic journals using an exponential function rather than a linear one highlights the idea that, on average, publications in highly ranked journals require much more scholarly effort and are significantly more time-consuming to prepare, revise, and resubmit. This is especially true because these journals have much more rigorous peer review processes and more demanding reviewers compared with lower-ranked journals. Highly ranked journals tend to be far more selective, with acceptance rates below 10%, than their lower-ranked counterparts.

**Full Counting Vs. Fractional Counting Approaches to Publishing Productivity**

Our sample includes STEMM scientists only where collaborative publications are the rule and solo research the exception (Olechnicka et al., 2019; Wagner, 2018). In the full counting methods, equal full credits go to all coauthors; in fractional counting methods, credits are divided by the number of coauthors (our sample does not include articles with more than 100 coauthors which are found mostly in subdisciplines of physics and astronomy; see Waltman & van Eck, 2019). Consequently, the four combinations of counting and journal prestige–normalization methods have been considered in examining productivity—two full counting methods and two fractional counting methods—leading to four productivity types being used in the present research: (1) Productivity 1 (prestige normalized, full counting), (2) Productivity 2 (prestige normalized, fractional counting), (3) Productivity 3 (non-normalized, full counting), and (4) Productivity 4 (non-normalized, fractional counting).

Productivity distribution by productivity type (Productivities 1 through 4), career stage (assistant professorship period, associate professorship period), and field of science is shown in Fig. 2: The steepest distribution for every field occurs for both prestige-normalized approaches to productivity and the most flat for non-normalized, full counting approaches.

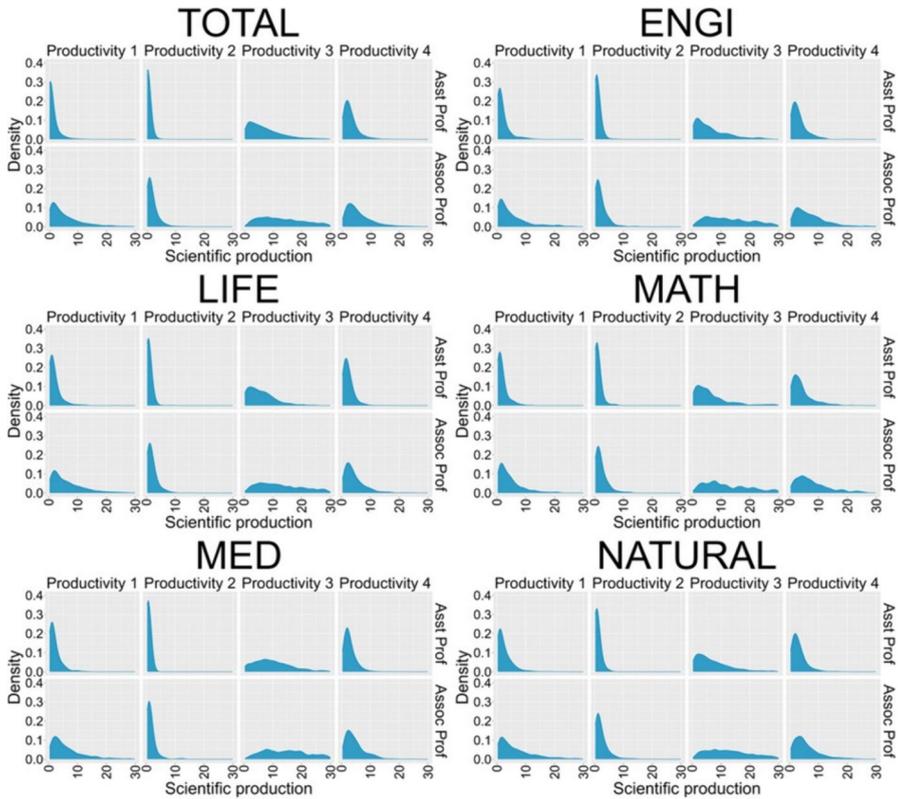

**Fig. 2** Kernel density plots, productivity distribution by productivity type (Productivities 1 through 4), career stage (assistant professorship period, associate professorship period) and field of science. Top panel (All fields of science combined; ENGI); Middle Panel (LIFE; MATH), and Bottom Panel (MED; NATURAL). Right-tails cut at 30 articles (N = 4,165)

### Allocating Scientists to Productivity Classes

A methodologically critical element of the present research is the allocation of scientists to the 10 (decile-based) productivity classes. First, separately within each field of science, all current associate professors are ranked according to their four-year productivity in the 2018–2021 reference period. There are four ranking procedures because there are four productivity types. The upper 10% of scientists within each field of science (the productivity decile 10 according to Productivities 1 through 4) are classified as the top productivity class, and the lowest 10% (the productivity decile 1) as the bottom productivity class, cut-off points permitting. There are 419 scientists in the top productivity class and 412 scientists in the bottom productivity class (413 in Productivity 4, see Table 3).

One approach to top and bottom performance classes is to follow the data (e.g., using productivity brackets) and cluster individuals above the minimum threshold (top classes) and below a minimum threshold (bottom classes); the approach we use better fits the comparative nature of our study, especially in the modeling section in which predictors changing the odds of membership in top and bottom classes can be juxtaposed. We have also tested an approach in which jumps (and drops) in productivity percentiles that are defined individually between the two career stages are examined, with success being high jumps. However, the idea of top performers and bottom performers would need to be replaced by that of (top) risers and (top) droppers; as a result, the productivity decile approach has been found to be more fruitful.

Decile productivity classes are merely statistical devices to deal with continuous, highly skewed productivity distribution (Fig. 2), and they do not refer to the lives of scientists; there are different roads to high productivity and different reasons to have low productivity (see Wang & Barabási, 2021: 13–15 on Shockley's model; Bornmann, 2024 on the Anna Karenina principle). Publishing productivity is a single dimension of research performance that also includes such features as individual citation impact and citation impact per publication, publications in the top 10% journals, top 10% cited publications, research grants awarded and their prestige, keynote speeches given at major conferences, doctoral students supervised, and so forth. Publishing productivity is a single aspect of research activities that, in turn, is a single aspect of wider university-based academic activities that traditionally also include teaching and service.

We have studied in detail the data in which the current decile-based classification is founded: For each field, we have examined the cut-off points between deciles of productivity, from bottom to top (Table 2, Productivity 1 only). An analysis performed separately for all fields of science and the two career stages shows powerful differences in cut-off points within fields between scientists in the career stages of assistant professorship and associate professorships for both cut-off points for decile 1 and decile 10. For instance, for the cut-off points between productivity decile 9 and decile 10, the differences are in the range of 4–6 times (from on average 4.73 in MATH to on average 5.98 in ENGI with 5.33 for all fields combined), implying that associate professors publish much more and in much more highly ranked journals. We have also examined productivity distribution (visualized using Kernel density plots) by productivity type and career stage; as expected, the distribution for the two prestige-normalized productivity types are much steeper than for the two non-normalized productivity types; and in all fields combined, the distribution is much steeper for the assistant professorship than for associate professorship period (Fig. 2).

Next, again separately within each field of science and separately by the four productivity types, all current associate professors have been ranked according to their average four-year productivity when they were assistant professors. The examined period differs because there are different lengths of working at the rank of assistant professorship; however, the rankings are based on productivity in a four-year equivalent period.

Table 2 Cut-off points (publication numbers: articles and chapters in conference proceedings) for membership in productivity deciles, by career stage (assistant professorship and associate professorship), field of science, Productivity 1 (the data for Productivities 2 through 4 available upon request) (N = 4,165)

| Decile | ENGI | | LIFE | | MATH | | MED | | NATURAL | | Total | |
|---|---|---|---|---|---|---|---|---|---|---|---|---|
| | Asst Prof stage | Assoc Prof stage | Asst Prof stage | Assoc Prof stage | Asst Prof stage | Assoc Prof stage | Asst Prof stage | Assoc Prof stage | Asst Prof stage | Assoc Prof stage | Asst Prof stage | Assoc Prof stage |
| Min | 0.001 | 0.004 | 0.001 | 0.002 | 0.001 | 0.004 | 0.001 | 0.010 | 0.001 | 0.002 | 0.001 | 0.002 |
| 1 | 0.003 | 0.288 | 0.003 | 0.553 | 0.003 | 0.350 | 0.010 | 0.859 | 0.004 | 0.522 | 0.004 | 0.471 |
| 2 | 0.007 | 0.734 | 0.008 | 1.433 | 0.007 | 0.880 | 0.019 | 1.690 | 0.009 | 1.278 | 0.009 | 1.117 |
| 3 | 0.013 | 1.288 | 0.016 | 2.313 | 0.012 | 1.363 | 0.028 | 2.480 | 0.017 | 2.070 | 0.017 | 1.903 |
| 4 | 0.032 | 2.036 | 0.032 | 3.148 | 0.020 | 1.971 | 0.054 | 3.485 | 0.040 | 3.276 | 0.037 | 2.778 |
| 5 | 0.155 | 2.909 | 0.240 | 4.560 | 0.081 | 2.681 | 0.309 | 4.476 | 0.205 | 4.665 | 0.195 | 3.899 |
| 6 | 0.395 | 4.283 | 0.498 | 6.097 | 0.304 | 3.753 | 0.667 | 5.952 | 0.568 | 6.480 | 0.478 | 5.422 |
| 7 | 0.731 | 5.973 | 0.956 | 8.035 | 0.634 | 5.003 | 1.179 | 7.835 | 1.089 | 8.867 | 0.906 | 7.476 |
| 8 | 1.252 | 8.610 | 1.535 | 10.975 | 1.222 | 6.755 | 1.732 | 10.771 | 2.110 | 12.603 | 1.615 | 10.393 |
| 9 | 2.510 | 15.010 | 2.738 | 15.777 | 2.129 | 10.063 | 3.140 | 16.269 | 4.008 | 19.758 | 3.055 | 16.271 |
| Max | 32.430 | 81.313 | 23.640 | 76.806 | 11.883 | 39.775 | 14.668 | 116.741 | 308.016 | 762.767 | 308.016 | 762.767 |

**Table 3** How the current top-performing (productivity decile 10) associate professors (Left panel) and the current bottom-performing (productivity decile 1) associate professors (Right panel) are distributed by productivity percentiles (range: 0–100) when they were assistant professors. Associate professors, initial (as assistant professors) percentile distribution statistics, Productivity 1: full counting prestige-normalized, by academic field, gender, type of institutional research intensity, academic age group, and biological age group (N=4,165)

|  |  | Current top-performing (top 10% productivity) associate professors | | | | Current bottom-performing (bottom 10% productivity) associate professors | | | |
| --- | --- | --- | --- | --- | --- | --- | --- | --- | --- |
|  |  | N | Mean | Std dev | Median | N | Mean | Std dev | Median |
| Total |  | N=419 | 78.6 | 22.2 | **87.9** | N=412 | 26.1 | 23.2 | **18.3** |
| Field | ENGI | N=96 | 78.5 | 23.4 | **88.2** | N=95 | 33.0 | 24.8 | **29.6** |
|  | LIFE | N=90 | 76.3 | 23.3 | **84.7** | N=89 | 23.3 | 22.0 | **17.2** |
|  | MATH | N=41 | 77.3 | 21.9 | **86.3** | N=39 | 21.2 | 20.7 | **13.6** |
|  | MED | N=64 | 78.4 | 21.0 | **86.6** | N=62 | 22.7 | 22.1 | **15.5** |
|  | NATURAL | N=128 | 80.8 | 21.2 | **89.8** | N=127 | 25.9 | 23.2 | **18.0** |
| Gender | Female scientists | N=140 | 78.5 | 21.8 | **88.5** | N=1`33 | 28.0 | 24.8 | **18.7** |
|  | Male scientists | N=279 | 78.6 | 22.4 | **87.6** | N=279 | 25.1 | 22.4 | **18.1** |
| Institutional research intensity | IDUB | N=161 | 81.2 | 21.7 | **91.3** | N=91 | 29.3 | 24.7 | **21.2** |
|  | Rest | N=258 | 76.9 | 22.4 | **85.2** | N=321 | 25.1 | 22.7 | **18.0** |
| Academic age groups | Beginning | N=6 | 78.5 | 16.2 | **80.7** | N=7 | 40.1 | 21.5 | **44.6** |
|  | Early | N=201 | 89.0 | 14.2 | **93.7** | N=128 | 31.8 | 26.5 | **30.0** |
|  | Middle | N=198 | 69.2 | 24.0 | **74.5** | N=195 | 25.1 | 23.1 | **16.5** |
|  | Late | N=14 | 60.9 | 23.2 | **55.4** | N=82 | 18.0 | 13.3 | **16.2** |
| Biological age groups | 39 and less | N=104 | 92.7 | 10.9 | **96.9** | N=5 | 72.9 | 14.3 | **68.6** |
|  | 40–54 | N=289 | 76.2 | 21.9 | **84.3** | N=148 | 39.3 | 25.1 | **42.8** |
|  | 55 and more | N=26 | 48.7 | 20.3 | **42.2** | N=259 | 17.6 | 16.7 | **12.0** |

## Results

### Mobility Patterns Between Productivity Classes

Our focus is on the mobility between productivity classes (especially between top and bottom classes and the classes closest to them: productivity deciles 8, 9, and 10 at the top; and productivity deciles 1, 2, and 3 at the bottom). Assistant professors from the top and bottom productivity classes can change their productivity classes while being associate professors, moving to top, bottom, or any other productivity decile. Specifically, we analyze the following mobility types by field of science and productivity type:

(1) *Top-to-top mobility* (assistant professors belonging to the top productivity class continue to belong to the top productivity class as associate professors: mobility from productivity decile 10 to decile 10);

(2) *Bottom-to-bottom mobility* (assistant professors belonging to the bottom productivity class continue to belong to the bottom productivity class as associate professors: mobility from productivity decile 1 to decile 1);

(3) *Extreme downward and extreme upward mobility: top-to-bottom mobility* and *bottom-to-top mobility* (assistant professors belonging to the top productivity class move down to the bottom productivity class as associate professors; and, analogously, assistant professors belonging to the bottom productivity class move up to the top productivity class as associate professors; mobility from productivity decile 10 to decile 1; mobility from productivity decile 1 to decile 10).

Apart from the above basic mobility types that include only productivity deciles 10 and 1, we will also discuss briefly a wider mobility between the upper productivity deciles (8–10) and the lower productivity deciles (1–3), here taking into consideration the role of cut-off points in publication numbers (Table 2). There are near-hit and near-miss observations in our datasets: Scientists just above the decile 1 cut-off point and just below the decile 10 cut-off point; hence, a more general view which includes neighboring deciles seems useful.

Our general question is how the current top-performing (productivity decile 10, $N = 419$) associate professors are distributed by productivity percentile ranks (range: 0–100) when they were assistant professors in the past. In addition, analogously, we are interested in how the current bottom-performing (productivity decile 1, $N = 412$) associate professors are distributed by productivity percentile ranks (range: 0–100) when they were assistant professors in the past.

We have examined these questions using distributions by field of science, gender, institutional research intensity, and two age-related variables (academic age group and biological age group). As could be expected, the median value of the original percentile rank (as assistant professor) for the current top performers is very close to the target percentile rank (as associate professor): The median is 87.9 percentile for top performers and 18.3 percentile for bottom performers (Table 3), with limited field-related variability for top performers (from 84.7 in LIFE to 89.8 in NATURAL) and considerable field-related variability for bottom performers (from 13.6 in MATH to 29.6 in ENGI). Gender differences are marginal: Currently top-performing (and bottom-performing) men and women as associate professors on average come from similar initial productivity percentile ranks as assistant professors.

An instructive way to visualize an answer to the questions of how current top-performing (and bottom-performing) associate professors are located in terms of productivity deciles when they were assistant professors is to use Kernel density plots (Fig. 3). Kernel density plots use kernel density estimation to create a smoothed, continuous curve that approximates the underlying data distribution. These estimations work better than histograms in displaying the shape of a distribution because their shape is not affected by the number of bins used or by dramatic differences between them; they can be flexibly used to compare distributions of two or more datasets. For all fields combined, the vast majority of current top performers belonged to productivity deciles 8 through 10 in the past, and the vast majority of bottom performers belonged to productivity deciles 1 through 3 in the past. However, the highest concentration of top-performing associate professors is observed for NATURAL and of bottom-performing scientists for MATH. The highest concentration of top performers is observed for the youngest age group (scientists aged 39 and less, Fig. 3).

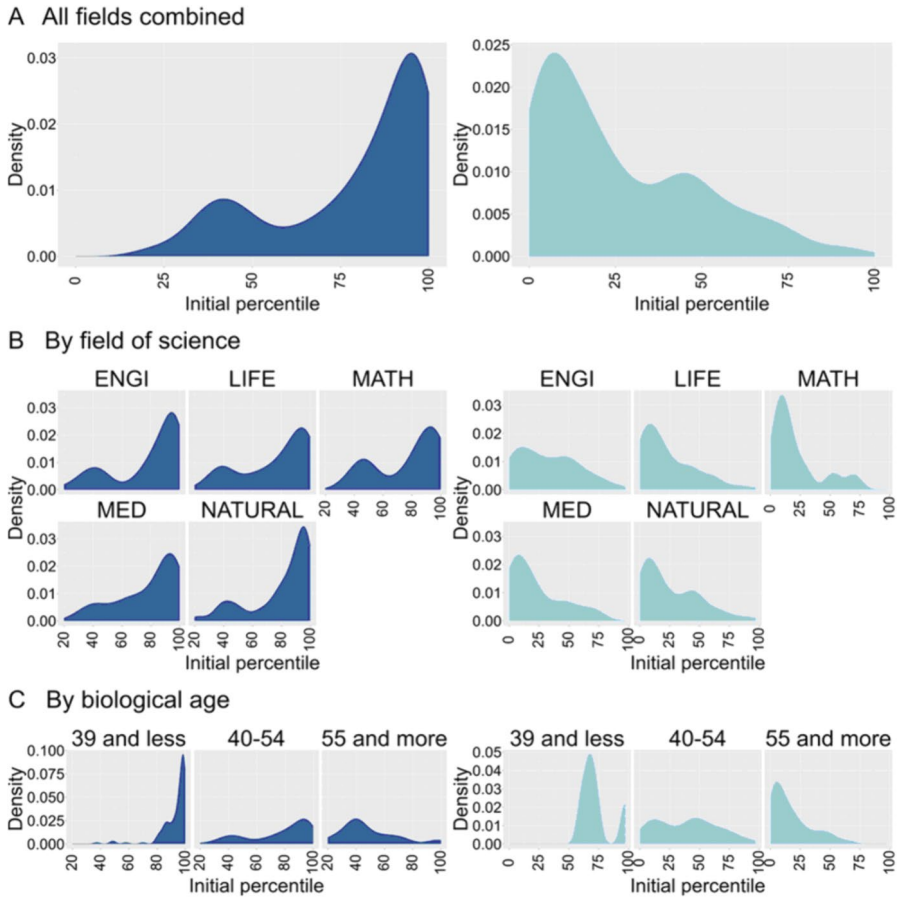

**Fig. 3** How the current top-performing (Left, N=419, productivity decile 10) and bottom-performing (Right, N=412, productivity decile 1) associate professors are distributed by productivity percentiles (range: 0–100) when they were assistant professors. **A**. all fields combined, **B**. by field of science. **C**. by biological age groups. Kernel density plots, initial percentile distribution, Productivity 1: full counting, prestige-normalized

With our dataset, we can analyze the mobility between productivity deciles (at the level of individuals) in great detail. Table 4 (Top panel) shows the decile origins of scientists currently located in productivity decile 10: initial productivity deciles (as assistant professors) in the past of current top-performing associate professors across the various academic fields.

Almost half of the current highest-performing associate professors come from productivity decile 10 during their time as assistant professors (46.5%): They continue to be in the same productivity decile (17.7% come from decile 9 and 8.6% from decile 8). In total, three-fourths of them belonged to productivity deciles 8–10 in the past (72.8%). Almost none belonged to the lowest three productivity deciles, with no scientist experiencing the extreme upward decile 1 to decile 10 mobility and just one scientist (located in NATURAL) experiencing the upward decile 2 to decile 10 mobility (we have full lifetime biographical and publishing profiles of every scientist, including this exception).

Considering the academic fields, half of the high-performing associate professors (50.0%) were also high-performing assistant professors in NATURAL and 47.9% in ENGI. In ENGI, 74.0% of top performers come from the highest three deciles, and none comes from the lowest three. In LIFE and MATH, these figures are 67.8% and 70.7%, 2.2%, and 0%, respectively.

Table 4 (Bottom panel) shows the decile origins of scientists currently located in productivity decile 1: initial productivity deciles (as assistant professors) in the past of current bottom-performing associate professors across the various academic fields. The emergent patterns are mirror-like but less intense than those found for top performers: About two-thirds (63.9%) of bottom performers come from the three lowest productivity deciles (deciles 1, 2 and 3), including one-third from the lowest decile (33.3%). Only 5.5% (23 individual scientists) come from the highest three deciles, and of these individuals, again, we have full lifetime biographical and publishing data.

We discuss mobility between productivity classes based on the four productivity types first (1) for all fields of science combined and, second, by (2) zooming in on cross-field differences: horizontal top-to-top mobility, horizontal bottom-to-bottom mobility, and extreme upward and downward mobility (bottom-to-top mobility, top-to-bottom mobility). Finally, the results section presents (3) the logistic regression analysis, which examines the predictors changing the odds of entering the classes of top productivity and bottom productivity associate professors.

### Mobility between Productivity Classes by Productivity Type: All Fields of Science Combined

The Sankey diagram (Fig. 4) provides a guiding visualization to better understand what is meant by scientists' mobility across productivity classes. The Sankey diagram shows flows of scientists between the productivity deciles of assistant professors (left: top and bottom) and associate professors (right: top and bottom): Specifically, horizontal top-to-top and bottom-to-bottom mobility, as well as downward top-to-bottom and upward bottom-to-top mobility, are of interest to us here.

The example in Fig. 4 shows the mobility of scientists from all fields of science combined and uses Productivity 1 type (prestige-normalized, full counting approach). The left column shows the distribution of assistant professors within top and bottom productivity classes (totaling 100% in each class), and the right column

Table 4 Mobility of top performers between productivity deciles between two career stages—assistant professorship (initial stage) and associate professorship (target stage). From which initial productivity deciles (as assistant professors) do current top-performing (Top panel) and bottom-performing (Bottom panel) associate professors come? Top-performing (N=419) and bottom-performing (N=412) associate professors by academic field and initial productivity decile, Productivity 1: full counting prestige-normalized (frequencies and percentages)

| | | Total | Bottom 10% | Decile 2 | Decile 3 | Decile 4 | Decile 5 | Decile 6 | Decile 7 | Decile 8 | Decile 9 | Top 10% |
|---|---|---|---|---|---|---|---|---|---|---|---|---|
| *Current top performing associate professors* | | | | | | | | | | | | |
| Total | N | 419 | 0 | 1 | 7 | 34 | 37 | 15 | 20 | 36 | 74 | 195 |
| | % | **100.0** | **0.0** | **0.2** | **1.7** | **8.1** | **8.8** | **3.6** | **4.8** | **8.6** | **17.7** | **46.5** |
| ENGI | N | 96 | 0 | 0 | 2 | 10 | 9 | 1 | 3 | 7 | 18 | 46 |
| | % | 100.0 | 0.0 | 0.0 | 2.1 | 10.4 | 9.4 | 1.0 | 3.1 | 7.3 | 18.8 | 47.9 |
| LIFE | N | 90 | 0 | 0 | 2 | 10 | 7 | 4 | 6 | 7 | 16 | 38 |
| | % | 100.0 | 0.0 | 0.0 | 2.2 | 11.1 | 7.8 | 4.4 | 6.7 | 7.8 | 17.8 | 42.2 |
| MATH | N | 41 | 0 | 0 | 0 | 2 | 9 | 1 | 0 | 4 | 6 | 19 |
| | % | 100.0 | 0.0 | 0.0 | 0.0 | 4.9 | 22.0 | 2.4 | 0.0 | 9.8 | 14.6 | 46.3 |
| MED | N | 64 | 0 | 0 | 1 | 4 | 4 | 4 | 5 | 7 | 11 | 28 |
| | % | 100.0 | 0.0 | 0.0 | 1.6 | 6.3 | 6.3 | 6.3 | 7.8 | 10.9 | 17.2 | 43.8 |
| NATURAL | N | 128 | 0 | 1 | 2 | 8 | 8 | 5 | 6 | 11 | 23 | 64 |
| | % | 100.0 | 0.0 | 0.8 | 1.6 | 6.3 | 6.3 | 3.9 | 4.7 | 8.6 | 18.0 | 50.0 |
| *Current bottom-performing associate professors* | | | | | | | | | | | | |
| Total | N | 412 | 137 | 81 | 45 | 30 | 49 | 24 | 23 | 15 | 3 | 5 |
| | % | **100.0** | **33.3** | **19.7** | **10.9** | **7.3** | **11.9** | **5.8** | **5.6** | **3.6** | **0.7** | **1.2** |
| ENGI | N | 95 | 22 | 14 | 12 | 10 | 14 | 9 | 5 | 6 | 1 | 2 |
| | % | 100.0 | 23.2 | 14.7 | 12.6 | 10.5 | 14.7 | 9.5 | 5.3 | 6.3 | 1.1 | 2.1 |
| LIFE | N | 89 | 32 | 19 | 11 | 6 | 10 | 2 | 6 | 1 | 1 | 1 |
| | % | 100.0 | 36.0 | 21.3 | 12.4 | 6.7 | 11.2 | 2.2 | 6.7 | 1.1 | 1.1 | 1.1 |
| MATH | N | 39 | 14 | 12 | 4 | 2 | 0 | 4 | 1 | 2 | 0 | 0 |
| | % | 100.0 | 35.9 | 30.8 | 10.3 | 5.1 | 0.0 | 10.3 | 2.6 | 5.1 | 0.0 | 0.0 |

Table 4 (continued)

| | | Total | Bottom 10% | Decile 2 | Decile 3 | Decile 4 | Decile 5 | Decile 6 | Decile 7 | Decile 8 | Decile 9 | Top 10% |
|---|---|---|---|---|---|---|---|---|---|---|---|---|
| MED | N | 62 | 25 | 13 | 6 | 2 | 6 | 4 | 3 | 3 | 0 | 0 |
| | % | 100.0 | 40.3 | 21.0 | 9.7 | 3.2 | 9.7 | 6.5 | 4.8 | 4.8 | 0.0 | 0.0 |
| NATURAL | N | 127 | 44 | 23 | 12 | 10 | 19 | 5 | 8 | 3 | 1 | 2 |
| | % | 100.0 | 34.6 | 18.1 | 9.4 | 7.9 | 15.0 | 3.9 | 6.3 | 2.4 | 0.8 | 1.6 |

shows the distribution of associate professors within the same two productivity classes.

The horizontal top-to-top and bottom-to-bottom mobility is represented by thick flows. Extreme vertical top-to-bottom mobility is rare and is represented as a thin downward flow; bottom-to-top flow is not shown because it does not occur at all: Only 1.2% of top productivity assistant professors (five scientists) land in the class of bottom productivity associate professors—and no bottom productivity assistant professor (zero scientists) lands in the class of top productivity associate professors.

From an aggregated view of all fields of science combined (Table 5 and Supplementary Table 1), the mobility patterns are unambiguous: About half (46.5%) of scientists allocated to the top productivity classes (Decile 10) stay in the same top class in their academic career, and about one-third (33.3%) of scientists allocated to the bottom productivity classes (Decile 1) stay in the same bottom class (for Productivity 1). There is an interesting locking-in mechanism in academic careers that deserves further scholarly attention—which is especially interesting because advancement in academic careers (in the specific Polish case) is related only to publications and publishing productivity, with a marginal role played by teaching and service university missions.

Analyzing the current biological age distribution of associate professors (Fig. 1) and their current distribution by age groups (Table 1), we can conclude that STEMM scientists are stuck in productivity classes for years, sometimes decades: Almost two-thirds of associate professors are aged 55 and older (30.9%); they became assistant professors when they were, on average, aged about 28–32. In a system in which full professorship is the crowning achievement of one's academic career and is available to few only, associate professors are scattered across all age groups. Consequently, our analyses span several decades of the academic careers of current associate professors.

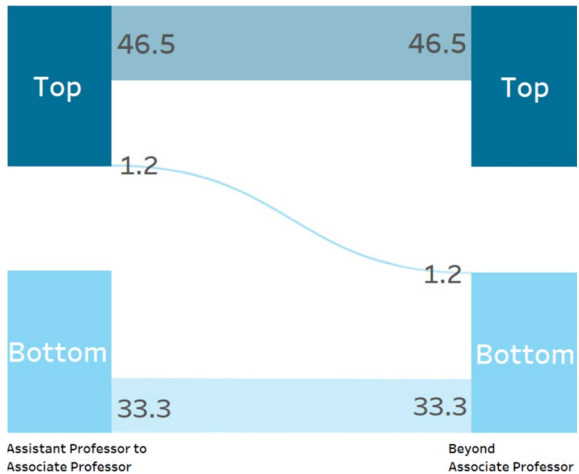

**Fig. 4** Example: Scientists' mobility between productivity classes in the two stages of an academic career. Productivity 1: prestige-normalized, full counting approach. All five STEMM fields of science combined, current associate professors. All observations ranked and clustered into productivity deciles, top (upper 10%, productivity decile 10, N=419) and bottom (bottom 10%, productivity decile 1, N=412) productivity classes only (percentages, top class and bottom class, 100% each)

Importantly, in our approach, we do not refer to publication numbers because productivity in Poland has been on the rise across the board over the past decade. We rank all current associate professors in terms of productivity ("target academic position" in productivity mobility in Table 5) and allocate them to productivity classes, separately for each field of science and separately by each productivity type. Subsequently, we rank associate professors retrospectively—that is, when they were assistant professors—in terms of productivity at that time, as measured for a four-year period ("initial academic position" in productivity mobility in Table 5). We examine 4,165 individual academic trajectories across five STEMM fields using full biographical, administrative, and bibliometric data at the micro-level of individual scientists.

What are the chances for extreme upward (decile 1 to decile 10) or downward (decile 10 to decile 1) mobility between productivity classes? Can scientists radically change their publishing behavior (compared with their peers in fields of science)?

**Table 5** Mobility between top (decile 10) and bottom (decile 1) productivity classes while moving up from the assistant professorship stage to associate professorship stage by four productivity types, fields of science combined (frequencies and percentages) (N=4,165)

| Assistant professorship stage (transition from) | Assistant professorship class (initial academic position) | Associate professorship stage (transition to) | Associate professorship class (target academic position) | Number of scientists in transition | Number of scientists in productivity class | % |
|---|---|---|---|---|---|---|
| *Productivity 1: Prestige-normalized full counting* | | | | | | |
| Asst Prof | Bottom | Assoc Prof | Bottom | 137 | 412 | 33.3 |
| Asst Prof | Bottom | Assoc Prof | Top | 0 | 412 | 0 |
| Asst Prof | Top | Assoc Prof | Bottom | 5 | 419 | 1.2 |
| Asst Prof | Top | Assoc Prof | Top | 195 | 419 | 46.5 |
| *Productivity 2: Prestige-normalized fractional counting* | | | | | | |
| Asst Prof | Bottom | Assoc Prof | Bottom | 122 | 412 | 29.6 |
| Asst Prof | Bottom | Assoc Prof | Top | 0 | 412 | 0 |
| Asst Prof | Top | Assoc Prof | Bottom | 5 | 419 | 1.2 |
| Asst Prof | Top | Assoc Prof | Top | 168 | 419 | 40.1 |
| *Productivity 3: Non-normalized full counting* | | | | | | |
| Asst Prof | Bottom | Assoc Prof | Bottom | 135 | 412 | 32.8 |
| Asst Prof | Bottom | Assoc Prof | Top | 1 | 412 | 0.2 |
| Asst Prof | Top | Assoc Prof | Bottom | 5 | 419 | 1.2 |
| Asst Prof | Top | Assoc Prof | Top | 191 | 419 | 45.6 |
| *Productivity 4: Non-normalized fractional counting* | | | | | | |
| Asst Prof | Bottom | Assoc Prof | Bottom | 111 | 413 | 26.9 |
| Asst Prof | Bottom | Assoc Prof | Top | 1 | 413 | 0.2 |
| Asst Prof | Top | Assoc Prof | Bottom | 4 | 419 | 1 |
| Asst Prof | Top | Assoc Prof | Top | 175 | 419 | 41.8 |

Our data (Table 5) clearly show that, in prestige-normalized counting approaches, there are no chances for extreme upward mobility: None of the current 412 top performers in the second career stage was a bottom performer in their first career stage (0%). In addition, a traditional non-normalized full counting approach shows merely one scientist (0.24%) experiencing this mobility. The chances for extreme downward mobility are slightly higher but still marginal—around 1% (5 scientists out of 419 or 1.19% in three productivity types and four scientists or 0.95% in Productivity 4).

**Zooming on Cross-Field Differences**

The aggregated picture of all fields of science combined hides a much more nuanced picture of individual STEMM fields, with their distinct mobility patterns between productivity classes. Focusing on Productivity 1 and the horizontal top-to-top (productivity decile 10 to decile 10) mobility first, for all fields, 40–50% of assistant professors continue in top productivity classes as associate professors (Table 6). The highest share is observed for natural sciences (NATURAL), with as much as 50.0%, followed by engineering (ENGI) with 47.9%. The lowest share is observed for life sciences (LIFE), where 42.2% of assistant professors continue in the same top productivity class. High cross-field differentiation is also observed for top-to-top mobility between productivity classes by the other three productivity types (see Supplementary Table 1).

Similarly, we have analyzed horizontal bottom-to-bottom mobility between productivity classes by field of science and productivity type. For all fields of science combined, the bottom-to-bottom mobility is experienced by one-third (33.3%) of current bottom-productive associate professors, and it is the highest when Productivity 1 is used and the lowest when Productivity 4 is used (33.3% and 26.9%, respectively). However, the differentiation by field of science within and across productivity types is considerably higher than in the case of top-to-top mobility (Supplementary Table 2). In ENGI, using Productivity 1, the percentage of scientists from the bottom productivity class as assistant professors staying in the same bottom productivity class as associate professors reaches 22.4%; in MED, it is twice as high (40.3%). Overall, between 20 and 40% of bottom performers in the first stage of their careers continue to be bottom performers in the second stage.

Transitions from the top to the bottom decile are extremely rare, with the highest chances being 2.1% in engineering and 1.6% in natural sciences. Overall, only 1.2% of assistant professors moved downwards from the top to the bottom decile. No such transitions occur in MATH and MED (0%).

Scientists representing bottom-to-top mobility, which is of great interest in productivity studies, are nonexistent across all fields of science in prestige-normalized productivity types (0%, Table 7). The only exception is a single scientist in life sciences (1.1% in LIFE) in each of the two non-normalized productivity types, hence totaling two scientists. By way of example of the power of microdata in our Laboratory of Polish Science, these two outliers are a male and a female scientists working in the discipline of AGRI in non-research-intensive institutions, aged 44 and 48, with doctoral degrees both aged 30 and habilitation degrees aged 41 and 39; both publish in relatively lower-ranked Scopus journals (median journal prestige percentile rank, lifetime: 32 and 25); their average team sizes are relatively high for AGRI (5.21 and 4.77 authors per paper, lifetime); their total research output lifetime is 14 and 40 articles; and they jumped from the 6th and 4th percentile in productivity distribution to the 90th percentile, respectively. We can compute a dozen other individual-level data about their academic careers (as well as about any other's), including their individual field-normalized citation-based impact, the citation-based impact of each of their papers, their promotion speed and promotion age compared with peers in AGRI from the same age cohort, publishing patterns and collaboration patterns changing over time, and so forth.

Table 6 Four mobility types by field of science (Productivity 1, prestige-normalized full counting), percentages (N = 4,165)

| Field of science | Top-to-top mobility | | Bottom-to-bottom mobility | | Top-to-bottom mobility | | Bottom-to-top mobility | |
|---|---|---|---|---|---|---|---|---|
| | Assistant professors: top to top (%) | As % of top associate professors | Assistant professors: bottom to bottom (%) | As % of bottom associate professors | Assistant professors: top to bottom (%) | As % of bottom associate professors | Assistant professors: bottom to top (%) | As % of top associate professors |
| ENGI | 47.9 | 47.9 | 22.4 | 23.2 | 2.1 | 2.1 | 0 | 0 |
| LIFE | 42.2 | 42.2 | 36.0 | 36.0 | 1.1 | 1.1 | 0 | 0 |
| MATH | 46.3 | 46.3 | 36.8 | 35.9 | 0 | 0 | 0 | 0 |
| MED | 43.8 | 43.8 | 40.3 | 40.3 | 0 | 0 | 0 | 0 |
| NATURAL | 50.0 | 50.0 | 35.2 | 34.6 | 1.6 | 1.6 | 0 | 0 |
| Total | 46.5 | 46.5 | 33.3 | 33.3 | 1.2 | 1.2 | 0 | 0 |

## Model Approach: Logistic Regression

First, we discuss four models—one for each productivity type—by estimating the odds ratios of membership in the top productivity class for associate professors (top 10%); and second, we do the same for membership in the bottom productivity class (bottom 10%). Each model evaluates the influence of several variables on the likelihood of being a top-productive (or a bottom-productive) associate professor.

### Logistic Regression: Top Productivity Associate Professors

Four logistic regression models have been constructed for four productivity types, where success is entering the class of the 10% most productive associate professors. The selection of variables is guided by the literature on productivity (e.g., Lee & Bozeman, 2005; Ramsden, 1994; Shin & Cummings, 2010; Teodorescu, 2000) and high productivity (e.g., Abramo et al., 2009a, 2009b; Fox & Nikivincze, 2021) as well as by data availability. An analysis of the presence of collinearity among the independent variables is performed. For this purpose, inverse correlation matrices are estimated, and the values from their main diagonals are used (see Supplementary Table 4).

One predictor proves to be the most important in the four models used: associate professors' membership in the class of top productivity assistant professors earlier in their careers (we track exactly the same scientists changing productivity classes over time, variable: Top_assistant_class, Table 8). In Model 1, this increases the odds by nearly sixfold (Exp(B) = 5.978, 95% CI: 4.493–7.954). Similar strong effects are seen in the other models, with odds ratios of 6.735 in Model 3, 3.677 in Model 2, and 6.305 in Model 4. In all four cases, the significance level is lower than 0.001.

Thus, the multidimensional analysis strongly confirms the results of the two-dimensional analysis presented in previous sections: Given the joint effect of all variables, the impact of membership in top productivity class in the past as an assistant professor (all other things being equal) is by far the strongest predictor of the current membership in top productivity class.

**Table 7** Bottom-to-top mobility by academic field and productivity type (percentages) (N = 4,165)

| Field of science | Productivity 1. Prestige-normalized full counting | | Productivity 2. Prestige-normalized fractional counting | | Productivity 3. Non-normalized full counting | | Productivity 4. Non-normalized fractional counting | |
|---|---|---|---|---|---|---|---|---|
| | Assistant professors: bottom to top (%) | As % of top associate professors | Assistant professors: bottom to top (%) | As % of top associate professors | Assistant professors: bottom to top (%) | As % of top associate professors | Assistant professors: bottom to top (%) | As % of top associate professors |
| ENGI | 0.0 | 0.0 | 0.0 | 0.0 | 0.0 | 0.0 | 0.0 | 0.0 |
| LIFE | 0.0 | 0.0 | 0.0 | 0.0 | 1.1 | 1.1 | 1.1 | 1.1 |
| MATH | 0.0 | 0.0 | 0.0 | 0.0 | 0.0 | 0.0 | 0.0 | 0.0 |
| MED | 0.0 | 0.0 | 0.0 | 0.0 | 0.0 | 0.0 | 0.0 | 0.0 |
| NATURAL | 0.0 | 0.0 | 0.0 | 0.0 | 0.0 | 0.0 | 0.0 | 0.0 |
| Total | 0.0 | 0.0 | 0.0 | 0.0 | 0.2 | 0.2 | 0.2 | 0.2 |

The models confirm our exploratory intuitions about changing productivity classes between the two career stages from a longitudinal perspective: When scientists are currently top productive at the stage of associate professorship, they tend to have been top productive at the earlier stage of assistant professorship. No matter how productivity is measured—full counting or fractional counting, prestige-normalized or non-normalized approaches—the patterns emerging from our regression analysis are very similar.

Prior membership in the promotion speed class of "fast" associate professors also proves to be statistically significant in all models. Fast associate professors are the scientists belonging to the 20% of scientists having the shortest time between their doctorate and habilitation, that is, between the start of their career as assistant professors and start of their career as associate professors. Belonging to a class of fast associate professors increases the probability of success: The speed at which one transitions to the associate professor level (variable: Fast_associate_class) increases the chances. In Model 1, the odds increase by 47.1% on average (Exp(B) = 1.471, 95% CI: 1.015–2.13; all other things being equal). This effect is even stronger in Model 4, where the odds more than double (Exp(B) = 2.128, 95% CI: 1.479–3.062; see Electronic Supplementary Material on promotion age and promotion speed classes). Rapid progression in an academic career appears to be a significant factor in achieving high productivity. Membership in the class of scientists receiving a doctorate at a young age (the top 20% of the distribution, variable: Young_assistant_class) is not statistically significant. In addition, institutional research intensity is not statistically significant.

Interestingly, gender appears in two models only, both times when using fractional counting. In Model 2, being male increases the odds of being a top-productive associate professor by 49% (Exp(B) = 1.49, 95% CI: 1.165–1.905) and in Model 4 by 33% (Exp(B) = 1.33, 95% CI: 1.042–1.697).

In three out of the four models (Models 1–3), biological age significantly and negatively affects the probability of success. In Model 1, each additional year of biological age decreases the odds by approximately 11% (Exp(B) = 0.888, 95% CI: 0.85–0.929). Similarly, in Models 2 and 3, the odds decrease by around 8–12% per year of age. This suggests that younger associate professors are more likely to be highly productive. Academic age, or the number of years since the first publication indexed in Scopus, in contrast, positively affects the odds in Models 1 and 2, both of which are prestige normalized. In Model 1, each additional year increases the odds by 8% (Exp(B) = 1.079, 95% CI: 1.049–1.11) and in Model 2 by 5.8% (Exp(B) = 1.058, 95% CI: 1.029–1.088).

Importantly, the direction and, to a large extent, strength of predictors generally do not depend on the model, that is, on the approach to productivity selected. Regardless of how productivity is measured, the statistically significant predictors are the same. By far, the strongest predictor of membership in the top productivity class of associate professors is prior membership in the top productivity class as assistant professor, which is consistent with our two-dimensional analyses of horizontal top-to-top mobility between productivity classes.

**Table 8** Logistic regression statistics: odds ratio estimates of membership in the class of top-productive associate professors (N=4,165)

| Model | Model 1: Prestige-normalized full counting | | | Model 2: Prestige-normalized fractional counting | | | Model 3: Non-normalized full counting | | | Model 4: Non-normalized fractional counting | | |
|---|---|---|---|---|---|---|---|---|---|---|---|---|
| | $R^2=0.22$, N=4165 | | | $R^2=0.21$, N=4165 | | | $R^2=0.20$, N=4165 | | | $R^2=0.18$, N=4165 | | |
| | Exp(B) | 95% C.I. for Exp(B) | | Exp(B) | 95% C.I. for Exp(B) | | Exp(B) | 95% C.I. for Exp(B) | | Exp(B) | 95% C.I. for Exp(B) | |
| | | Lower | Upper | | Lower | Upper | | Lower | Upper | | Lower | Upper |
| Male | 1.147 | 0.899 | 1.464 | 1.49** | 1.165 | 1.905 | 1.255 | 0.983 | 1.602 | 1.33* | 1.042 | 1.697 |
| Research intensive: Rest | 0.923 | 0.725 | 1.174 | 1.079 | 0.847 | 1.374 | 0.856 | 0.675 | 1.086 | 0.912 | 0.72 | 1.157 |
| Biological age | 0.888*** | 0.85 | 0.929 | 0.878*** | 0.839 | 0.918 | 0.922*** | 0.884 | 0.961 | 0.981 | 0.945 | 1.018 |
| Academic age | 1.079*** | 1.049 | 1.11 | 1.058*** | 1.029 | 1.088 | 1.025 | 0.998 | 1.052 | 0.988 | 0.964 | 1.012 |
| Assistant_age | 1.066 | 0.993 | 1.145 | 1.126** | 1.048 | 1.21 | 1.054 | 0.987 | 1.126 | 1.013 | 0.949 | 1.081 |
| Associate_age | 0.959 | 0.905 | 1.016 | 0.949 | 0.891 | 1.011 | 0.97 | 0.917 | 1.026 | 0.936 | 0.887 | 0.987 |
| Top_assistant_class | 5.978*** | 4.493 | 7.954 | 3.677*** | 2.776 | 4.871 | 6.735*** | 5.156 | 8.797 | 6.305*** | 4.85 | 8.197 |
| Young_assistant_class | 1.301 | 0.939 | 1.804 | 1.54 | 1.119 | 2.12 | 1.154 | 0.837 | 1.592 | 1.057 | 0.768 | 1.455 |
| Young_associate_class | 0.736 | 0.502 | 1.08 | 0.781 | 0.537 | 1.137 | 0.661* | 0.45 | 0.972 | 0.568** | 0.387 | 0.834 |
| Fast_associate_class | 1.471* | 1.015 | 2.13 | 1.63* | 1.129 | 2.354 | 1.825** | 1.265 | 2.633 | 2.128*** | 1.479 | 3.062 |
| Constant | 4.01 | 0.62 | 25.93 | 2.117 | 0.354 | 12.671 | 1.453 | 0.269 | 7.84 | 2.274 | 0.423 | 12.241 |

### Logistic Regression: Bottom Productivity Associate Professors

Subsequently, four logistic regression models have been built for four productivity types, where success is entering the class of 10% bottom-productive associate professors. Prior research on productivity and the productivity of Polish scientists in particular (e.g., Antonowicz et al., 2021; Kwiek, 2018) is instrumental in constructing the models. We discuss the results of standardized residuals statistics and the inverted correlation matrix main diagonal in the Electronic Supplementary Material.

The strongest predictor increasing the probability of membership in the class of bottom-productive associate professors is earlier membership in the class of bottom-productive assistant professors (Table 9). Being a bottom-productive assistant professor (the bottom 10% of the distribution, variable: Bottom_assistant_class) is a strong predictor across all models. In Model 1, this increases the odds by approximately 2.5 times (Exp(B) = 2.508, 95% CI: 1.882–3.342). Similar effects are seen in the other models, with odds ratios of 2.036 in Model 2, 3.566 in Model 3, and 2.929 in Model 4. Again, logistic regression analysis supports our two-dimensional results, hence showing the role of high horizontal bottom-to-bottom mobility, although not as strongly as in the case of top-productive associate professors.

Both biological and academic age are statistically significant predictors but in different directions. Biological age shows a consistent positive relationship with being in the bottom productivity decile across all models. For example, in Model 1, each additional year of biological age increases the odds by approximately 15.5% (Exp(B) = 1.155, 95% CI: 1.123–1.189). Similar effects can be observed in Models 2, 3, and 4, indicating that older associate professors are more likely to be less productive. Academic age, in contrast, negatively affects the odds in Models 1 and 2. In Model 1, each additional year decreases the odds by 4.1% (Exp(B) = 0.959, 95% CI: 0.942–0.976) and in Model 2 by 4% (Exp(B) = 0.960, 95% CI: 0.944–0.977).

The age at which one becomes an associate professor (variable: Associate_age) also shows some influence, with a negative effect in Models 1 and 2. For instance, in Model 1, each additional year decreases the odds by 5.9% (Exp(B) = 0.941, 95% CI: 0.903–0.981). This suggests that becoming an associate professor at a younger age slightly reduces the likelihood of low productivity.

The low speed at which one transitions to the associate professor level (variable: Slow_associate_class) shows a positive relationship with being in the bottom productivity decile in Model 4, where the odds increase by 51.1% (Exp(B) = 1.511, 95% CI: 1.048–2.178). This indicates that a slow transition might be associated with lower productivity.

Overall, the logistic regression models highlight several significant predictors for becoming a bottom-productive associate professor. Being a bottom-productive assistant professor is the most consistent and robust predictor, followed by biological age. Academic age and the promotion age for associate professorship also play important roles, with longer academic careers and younger promotion age reducing the odds of low productivity. Contrary to models for top-productive associate professors, gender was found to be non-significant. In addition, institutional research intensity is statistically significant in two models, increasing the odds by 37.9% in Model 1 and 30.7% in Model 3.

Table 9 Logistic regression statistics: odds ratio estimates of membership in the class of bottom-productive associate professors (N=4,165)

| Model | Model 1: Prestige-normalized full counting | | | Model 2: Prestige-normalized fractional counting | | | Model 3: Non-normalized full counting | | | Model 4: Non-normalized fractional counting | | |
|---|---|---|---|---|---|---|---|---|---|---|---|---|
| | $R^2=0.13$, N=4165 | | | $R^2=0.12$, N=4165 | | | $R^2=0.10$, N=4165 | | | $R^2=0.07$, N=4165 | | |
| | Exp(B) | 95% C.I. for Exp(B) | | Exp(B) | 95% C.I. for Exp(B) | | Exp(B) | 95% C.I. for Exp(B) | | Exp(B) | 95% C.I. for Exp(B) | |
| | | Lower | Upper | | Lower | Upper | | Lower | Upper | | Lower | Upper |
| Male | 1.110 | 0.879 | 1.403 | 1.135 | 0.904 | 1.424 | 1.033 | 0.821 | 1.300 | 1.016 | 0.814 | 1.270 |
| Research intensive: Rest | 1.379* | 1.065 | 1.786 | 1.102 | 0.864 | 1.405 | 1.307* | 1.012 | 1.686 | 1.035 | 0.815 | 1.315 |
| Biological age | 1.155*** | 1.123 | 1.189 | 1.062*** | 1.032 | 1.093 | 1.149*** | 1.117 | 1.182 | 1.038* | 1.008 | 1.068 |
| Academic age | 0.959*** | 0.942 | 0.976 | 0.984 | 0.968 | 1.001 | 0.96*** | 0.944 | 0.977 | 0.986 | 0.969 | 1.003 |
| Assistant_age | 0.993 | 0.941 | 1.047 | 0.950 | 0.901 | 1.001 | 0.991 | 0.941 | 1.045 | 0.965 | 0.915 | 1.018 |
| Associate_age | 0.941** | 0.903 | 0.981 | 1.047* | 1.006 | 1.089 | 0.952* | 0.914 | 0.993 | 1.038 | 0.997 | 1.081 |
| Bottom_assistant_class | 2.508*** | 1.882 | 3.342 | 3.566*** | 2.666 | 4.769 | 2.036*** | 1.523 | 2.723 | 2.929*** | 2.185 | 3.925 |
| Old_assistant_class | 0.815 | 0.548 | 1.213 | 0.743 | 0.503 | 1.098 | 0.738 | 0.499 | 1.091 | 0.727 | 0.491 | 1.075 |
| Old_associate_class | 1.120 | 0.782 | 1.605 | 1.316 | 0.925 | 1.873 | 1.319 | 0.929 | 1.872 | 1.314 | 0.921 | 1.875 |
| Slow_associate_class | 1.278 | 0.871 | 1.876 | 1.133 | 0.784 | 1.638 | 1.317 | 0.904 | 1.920 | 1.511* | 1.048 | 2.178 |
| Constant | 0.001*** | 0.000 | 0.007 | 0.003*** | 0.001 | 0.013 | 0.001*** | 0.000 | 0.006 | 0.008*** | 0.002 | 0.041 |

Generally, the predictors in those models constructed for bottom productivity associate professors are weaker than the predictors in the models constructed for top productivity associate professors. The directions of impact are usually opposite, which is especially visible in the case of age-related predictors. In both cases, the most powerful predictor in all models, that is, regardless of the productivity type used, is prior membership in the top productivity (or bottom productivity) class at the stage of assistant professorship.

## Discussion and Conclusion

The strength of the present research comes from its unique national datasets and solid methodological approach. First, this research has strong empirical foundations: The publishing productivity of all Polish internationally visible associate professors in 12 STEMM fields of science clustered into five fields of science (N=4,165) are studied, and their full individual publication portfolios and full individual biographical histories are examined. Biographical and demographic data from a national registry of scientists are combined with publication metadata, including all Polish articles published in the past half a century and indexed in Scopus (1973–2021, N=935,167).

Second, in terms of methodology, the individual scientist, rather than the individual publication, has been used as a unit of analysis; productivity based on the four major types is used to assess the extent to which measurement methods can impact the patterns found; and decile-based productivity classes rather than publication numbers are used. Specifically, journal prestige–normalized productivity (in which articles' locations in the global stratified journal system are applied) is contrasted with non-normalized productivity (both in full counting and fractional counting types).

Finally, a longitudinal research design has been used, in which publishing productivity over the years and decades, as linked to two subsequent career stages of assistant professorship and associate professorship, can be traced. Individuals are followed over time with full promotion, demographic, and (Scopus-indexed) publication data.

Our dataset includes publishing patterns of associate professors spanning across decades because, in the Polish system, there are no externally imposed time requirements for promotions: The associate professors in our sample work for decades in the system, and about one-third of them (30.9%) are aged 55 or older.

Our longitudinal analyses have focused on the mobility patterns between productivity classes for assistant and associate professors. In our leading productivity type (prestige-normalized, full counting), our results show that slightly less than a half (46.5%) of top productivity scientists (the top 10% of the distribution, productivity decile 10) continue as top productivity scientists; about one-third (33.3%) of bottom productivity scientists (the bottom 10% of the distribution productivity, decile 1) continue as bottom productivity scientists.

Extreme interclass mobility (downward top-to-bottom mobility, upward bottom-to-top mobility) has emerged in our research as a marginal phenomenon: 0–1.2% of all current associate professors, depending on productivity type, experience either type of this mobility in their career histories. They are extremely rare scholarly species: In a sample of 4,165 scientists, there are no scientists (0%) moving from bottom to top deciles in prestige-normalized productivity types (and merely two scientists in both non-normalized ones) and just five scientists moving from top to bottom deciles. Top-performing associate professors were predominantly top-performing assistant professors in the past (median productivity percentiles: 87.9), and bottom-performing associate professors were predominantly bottom-performing assistant professors in the past (median productivity percentiles: 18.3).

Both for scientists and decision makers in science policy at various levels, the message is challenging: Our analyses show that that radical changing publishing productivity levels (upward or downward: here the mobility between the top 10% and bottom 10% of the productivity distribution) in STEMM actually almost never happens in practice. In other words, some scientists tend to be highly productive for years and decades, and others—their colleagues in institutions and their peers within fields of science —tend to be bottom productive for years. There is a zero probability that scientists will be radically more and a marginal probability that they will be radically less productive when they move up the academic ladder.

Importantly, the cross-disciplinary differentiation is notable: In the natural sciences, the percentage of scientists experiencing top-to-top mobility reaches 50%. In addition, the aggregated picture at the level of fields of science of 33.3% of scientists from the bottom productivity class as assistant professors staying in the same bottom productivity class as associate professors hides a much more differentiated picture. Depending on the field, between 30 and 60% of bottom-productive assistant professors continue their careers as bottom-productive associate professors. Compared with the persistence of "stardom" examined by Abramo et al. (2017) for Italy, the shares of Polish top performers mainatining their high productivity in both career stages are higher (46.5.0% for all STEMM fields combined) than the shares of top scientists maintaining their high productivity or "stardom" in Italy over a period of 12 years (35%). Productivity stratification seems deeper and more long-lasting in Poland than in Italy, possibly because of decades of severe research underfunding.

Logistic regression analysis powerfully supports our two-dimensional results. In the case of the odds ratio estimates of the membership in the top productivity class of associate professors, one predictor proves to be the most important in the four models: the membership in the class of top-productive assistant professors earlier in their careers. This prior membership is statistically significant in all models to a similarly high degree, increasing the odds 4–6 times, depending on productivity type. Prior membership in the promotion speed class of fast associate professors is also statistically significant in all models. Membership increases the chances of success by 50%–130%, depending on the model. In the case of the odds ratio estimates of the membership in the bottom productivity class of associate professors, the strongest predictor is prior membership in the class of bottom-productive assistant professors. The probability of success increases by 150–300%, depending on the productivity type.

From a broader perspective, our study shows that the traditional notion of "strong" and "weak" individual track record in research—here linked only to publishing productivity and not to field-weighted citation impact, generation of research funding, or other dimensions of academic careers—makes sense when evaluating individual scientists.

Scientists with a very weak past track record in research emerge from our research as having marginal chances of becoming scientists with a very strong future track record across all STEMM fields. Traditionally, the science edifice has been constructed so that highly productive scientists—those with significant impact on disciplinary scientific communities—are highly recognized.

The takeaway from the present research for academic careers is that early achievements in science—when viewed through a proxy of early top publishing productivity—significantly influence achievements at a later stage, that is, late top publishing productivity. If the assistant professorship period is strong in publications, the associate professorship period tends to be strong too; analogously, if this period is weak in publications, associate professorship also tends to be weak.

Our micro-level data powerfully show that scientists tend to be stuck in their publishing productivity classes within their fields of science for years and decades: Top performers tend to be top performers, and bottom performers tend to be bottom performers; the former becoming the latter or the latter becoming the former are extremely rare phenomena. As a result, going to the individual productivity limits early on in academic careers tends to pay off later on in careers; later productivity is strongly related to earlier productivity.

However, one point needs to be remembered: in this study, our approach to productivity is relative rather than nominal. Productivity of individuals clustered into classes is compared to productivity of other individuals, also clustered into classes. While change is central to our analysis, it is the change between classes (i.e., relative) rather than the change in publication numbers (i.e., nominal). Our interest is not in increasing (or decreasing) productivity, in terms of publication numbers, with age (or as scientists become older and move up the academic ladder). In this sense, we can only speculate whether or not our top performers are actually increasing their productivity over time, following the idea of intensification of academic publishing (Hermanowicz & Scheitle, 2023); our focus is on changing productivity classes – which is a different issue, unrelated to traditional discussions about "age and scientific achievement" (e.g., Wang and Barabàsi, 2021).

The uncovered transition patterns are understandable in light of traditional productivity theories (especially sacred spark theory, e.g., Allison & Stewart, 1974; Cole & Cole, 1973; Fox, 1983; and cumulative advantage theory, e.g., David, 1994; DiPrete & Eirich, 2006; Merton, 1968). However, the strength of the patterns found in our research across different fields and different productivity types is somehow unexpected; what is especially surprising is the very high persistence of membership in top productivity classes across academic careers and the zero (and sometimes marginal) chance of radically changing productivity classes while moving up the academic ladder.

Hence, our research shows a long-term character of careers in science, with publishing productivity (and possibly other working patterns) in an apprenticeship period of assistant professorship heavily influencing productivity in a more independent period of associate professorship. We can assume that working patterns, which are often based on the role models of academic supervisors, the strength of teaching or research orientation, weekly working time distribution (including time spent on research), writing habits, and collaboration habits, take years to form and tend to stay with individuals in their careers. Science takes a lot of time (weekly, monthly, yearly); additionally, the academic careers of current highly productive scientists take time to form: Years or decades of previous high productivity are needed, as our data show.

Our research has its limitations. We do not analyze the productivity classes of all assistant professors because our focus is on associate professors publishing in the most recent four-year period for which publication data are available (2018–2021) and, retrospectively, when they were assistant professors. This means that the longitudinal comparison over time cannot be conducted for those assistant professors who have left the national science system. Our study also does not pertain to those scientists who never earned their postdoctoral degrees (they are not in our sample).

Our dataset does not include environmental factors, which bear heavily on individual productivity—we are not able to study the "work climate" (Fox & Mohapatra, 2007), which is reported to be especially important for women in the STEMM disciplines (Branch, 2016), save for a single variable of research-intensive institutions selected for a national excellence program. In a similar manner, we do not have the data about academic attitudes and behaviors, which are routinely reported in academic profession surveys, or the data on work–life balance, household and parenting obligations, which have been extensively used in previous survey-based productivity studies (e.g., Kwiek, 2019).

It also needs to be kept in mind that, at an individual level, generally, although productivity can always be higher in the higher education and science system, it cannot be much lower or zero (meaning nonpublishers, nonperformers) for a long time because these unsuccessful scientists tend to leave the Polish higher education system. Therefore, our study of current associate professors does not include failures in science because unsuccessful scientists are not present in the national registry. From this perspective, all current associate professors in our sample are successes in science ("success bias"), and it is merely a statistical approach that allocates them to the top and bottom productivity classes. No matter how highly productive the scientists in the system are, the system can always be divided into 10 decile-based productivity classes, and there will always be bottom productivity classes (productivity decile 1) in each field of science (cut-off points permitting). Finally, we use internationally understandable notions of assistant and associate professorships, even though, in fact, we use the two Polish academic degrees (doctorate and habilitation).

Our study is confined to a single national science system. The generalizability of the results depends on the similarities and dissimilarities with other systems. Science systems have differently constructed career ladders, and they differ in their internal competitiveness, incentive structures, teaching and research mix, funding opportunities, the attractiveness of academic careers, and so forth. However, our ongoing research (Kwiek & Szymula, 2024b) involves a large-scale cross-national study of mobility between productivity classes in 38 OECD countries, with 320,564 late-career scientists tracked over time.

**Supplementary Information** The online version contains supplementary material available at https://doi.org/10.1007/s10755-024-09735-3.

**Acknowledgements** We are grateful to the hosts and audiences of Marek Kwiek's seminars about longitudinal and class-based approaches to research productivity at the University of Oxford (Simon Marginson, CGHE, Center for Global Higher Education, June 2022), Stanford University (John Ioannidis, METRICS, Meta-Research Innovation Center, June 2022), DZHW Berlin (Torger Möller, German Center for Higher Education Research and Science Studies, June 2023), and Leiden University (Ludo Waltman, CWTS, Centre for Science and Technology Studies, June 2023). We are extremely grateful to Lukasz Szymula from the CPPS Poznan Team for his support in data collection and analysis. We gratefully acknowledge the assistance of the International Center for the Studies of Research (ICSR) Lab, with particular gratitude to Kristy James and Alick Bird. Finally, we would like to thank the co-editor in chief, Dr. John M. Braxton and the two anonymous reviewers whose friendly criticism was extremely helpful in developing our ideas.

**Funding** The authors gratefully acknowledge the support provided by the NDS grant no. NdS/529032/2021/2021.

**Data Availability** The authors used data from Scopus, a proprietary scientometric database. For legal reasons, data from Scopus received through collaboration with the Elsevier's ICSR Lab cannot be made openly available.

## Declarations

**Conflicts of Interests** The authors state that there are no conflicts of interest.

**Professor Marek Kwiek** holds the UNESCO Chair in Institutional Research and Higher Education Policy at the AMU University of Poznan, Poland. His research areas are quantitative science studies and higher education research and policy. His focus is on research collaboration, productivity, and social stratification in science. His most recent invited seminars include Berkeley, Harvard, Stanford, Oxford, Beijing, Shanghai, Hiroshima, Hong Kong, Leiden, Oslo, and Paris, among others. His recent monograph is Changing European Academics: A Comparative Study of Social Stratification, Work Patterns and Research Productivity (Routledge, 2019). An expert for the European Commission, European Parliament, USAID, World Bank, OECD, and UNESCO. A coordinating editor for Higher Education (SpringerNature). An ordinary member of the European Academy of Sciences and Arts (EASA) in Salzburg and Academia Europaea in London.

**Dr. Wojciech Roszka** is an assistant professor in the Department of Statistics, the Institute of Informatics and Quantitative Economics, Poznan University of Economics and Business, Poland. His area of research interest and expertise includes integration of data from different sources with particular emphasis on probabilistic record linkage, statistical matching, data fusion and microsimulation modelling. He is a member of the Polish Statistical Society and works in the Center for Public Policy Studies of the AMU University of Poznan.